%% file: arxiv1V2.tex
\documentclass[10pt,fleqn]{article}

\pdfoutput=1 

\usepackage{latexsym}
\usepackage{amsbsy}
\usepackage{amsfonts}
\usepackage{amssymb}
\usepackage{amscd}
\usepackage{epsfig}
\usepackage{color}
\usepackage{cite}

\mathchardef\Gamma="0100
\mathchardef\Delta="0101
\mathchardef\Theta="0102
\mathchardef\Lambda="0103
\mathchardef\Xi="0104
\mathchardef\Pi="0105
\mathchardef\Sigma="0106
\mathchardef\Upsilon="0107
\mathchardef\Phi="0108
\mathchardef\Psi="0109
\mathchardef\Omega="010A

\def\I{{\rm i}}
\def\D{{\rm d}}
\def\E{{\rm e}}

\def\vec{\boldsymbol}

\def\cal{\mathcal}

\input{diagxy}

\begin{document}




\title{Effect of Diffraction on Wigner Distributions  of 
 Optical  Fields and how to Use It in Optical Resonator Theory.\\ I -- Stable Resonators and Gaussian Beams}

 \date{}

  \maketitle
\begin{center}


  \vskip -1cm

{
\renewcommand{\thefootnote}{}
{\bf   Pierre Pellat-Finet\footnote{pierre.pellat-finet@univ-ubs.fr,
  eric.fogret@univ-ubs.fr} and \'Eric
  Fogret}
}
\setcounter{footnote}{0}

\medskip
{\sl \small Laboratoire de Mathématiques de Bretagne Atlantique UMR CNRS 6205

Université de Bretagne Sud, B. P. 92116, 56321 Lorient cedex, France}
\end{center}



\vskip 1cm

\begin{center}

\begin{minipage}{12cm}
\hrulefill

 \smallskip
{\small
{\bf Abstract.} The first part of the paper is
devoted to diffraction phenomena that can be expressed by fractional
Fourier transforms whose orders are real numbers. According to a scalar theory, diffraction acts on the amplitude of the electric field as well as  on its spherical angular
spectrum, and  Wigner distributions can be defined on a  space-frequency phase-space. The phase space is equipped with an Euclidean structure,
so  that the effects of diffraction are 
rotations of Wigner distributions associated with optical
fields. Such a rotation is shown to split into two specific elliptical rotations. Wigner distributions associated with
transverse modes of a resonator are invariant in these rotations,
and a complete theory of stable optical resonators and Gaussian beams is developed on the basis of this
property, including waist existence and  related formulae, and
naturally introducing the Gouy phase.

\smallskip
\noindent {\sl Keywords:} Diffraction, Fourier optics,   fractional order Fourier
transformation,  Gaussian
beams, optical resonators, spherical angular
spectrum,  Wigner distribution.

\smallskip

\noindent{\sl PACS:} 42.30.Kq

\smallskip
\noindent {\bf Content}

\smallskip

\noindent 1. Introduction \dotfill \pageref{sect1}

\noindent 2. Field transfer by  diffraction: Real-order transfer \dotfill \pageref{sect2}

\noindent 3. Transfer of the spherical angular spectrum \dotfill \pageref{sect3}

\noindent 4. Wigner distribution associated with the field \dotfill \pageref{sect4}

\noindent 5. Effect of  diffraction on the Wigner distribution:  Real-order transfer \dotfill \pageref{sect5}

\noindent 6. Composition of two tranformations \dotfill \pageref{sect6}

\noindent \ref{sect7}. Application to optical resonator theory \dotfill \pageref{sect7}

\noindent \ref{sect8}. The field amplitude on a wave surface. Longitudinal modes \dotfill \pageref{sect8}

\noindent \ref{sect9}. Concluding graphical analysis \dotfill \pageref{sect9}

\noindent Appendix A\dotfill \pageref{appenA}

\noindent Appendix B\dotfill \pageref{appenB}

\noindent  References\dotfill \pageref{refe}

}

\hrulefill
\end{minipage}
\end{center}

\section{Introduction}\label{sect1}

A lot of works have been devoted to Wigner distributions in many
areas, such as Quantum Mechanics or Signal Processing \cite{Alo}. In
Optics, since the works of Walther \cite{Wal} and
Bastiaans \cite{Bas},  Wigner distributions have been used for representing optical
fields, for dealing with radiometry and coherence theories \cite{Wal}, with applications to
tomography \cite{Alo}, or more recently for simulating wave effects in graphics
\cite{Cuy} or in ray tracing \cite{Mou}.

If we restrict our attention to light propagation, it has been shown that
the effect of a GRIN medium is a rotation of 
 the Wigner distribution associated with the optical field
\cite{Men,Oza3}. The link has been made with real-order fractional 
Fourier transformations, whose effects are also  rotations in an appropriate
phase-space \cite{Alo,Men,Oza3,Loh,Coe1}.  
Nevertheless, the  effect of Fresnel diffraction or of
propagation in free space (through Fresnel transforms), considered
between two transverse planes, 
is generally seen like a ``horizontal'' shear of  the 
Wigner distribution \cite{Alo,Loh,Tes}, not a rotation.  We notice that
Lohmann expresses Fraunhofer diffraction as a $\pi/2$--rotation,
but does not generalize to Fresnel diffraction \cite{Loh}, so that the
previous descriptions introduce a breaking between the effects of
Fresnel or Fraunhofer phenomena, a shearing or a rotation.

In the following, we consider diffraction in a broad meaning, including both Fresnel and Fraunhofer phenomena. The difference between them can be made as follows: the integral expressing the field transfer by diffraction---see Eq. (\ref{eq1})---includes a quadratic phase factor when expressing a Fresnel phenomenon, and no quadratic phase factor for a Fraunhofer phenomenon. (The quadratic phase factor in front of the integral in Eq. (\ref{eq1}) does not matter, since it has no effect on the irradiance of the diffraction pattern.)

The effect of 
diffraction (scalar theory) on Wigner distributions associated
with optical fields can be obtained by linking two results:
\begin{enumerate}
\item A 
  diffraction phenomenon between a spherical emitter and a spherical receiver is expressed by a fractional order Fourier
  transform \cite{PPF1,PPF3,PPF5}.
\item Once  chosen appropriate scaled variables, fractional order Fourier transformations operate as ro\-ta\-tions in the
  phase space on which Wigner distributions are de\-fi\-n\-ed \cite{Alo,Alm,Men,Oza3,Loh,Coe1}.
\end{enumerate}

As
far as we know, no work has been dealing with establishing such a
link, which constitutes the main subject of the present paper,
and clearly, the result will be that the effect of
diffraction is a rotation of the Wigner distribution. We do not
obtain a
shearing, as proposed by several authors \cite{Alo,Loh,Tes}, because
these authors consider diffraction between two planes, and not between
spherical caps as we do.  Our approach
will make the above mentioned breaking between Fresnel
and Fraunhofer phenomena disappear: generally, Fraunhofer diffraction is
physically obtained from Fresnel diffraction by continuously
increasing the distance at which the diffracted irradiance is observed; 
its effect on the Wigner
representation will be deduced from the effect of Fresnel diffraction
 by continuously varying the rotation
angle up to $-\pi /2$. 

The optical field is described by a function of two real spatial variables, that is a function of a 2--dimensional vector variable, so that
the corresponding phase-space is 4--dimensional and the Wigner
distribution is a function of four real variables. Almost every paper on the
Wigner function deals with functions of  time (1--dimensional
variable); even when
dealing with optical fields, most authors (excepting Lohmann \cite{Loh}) restrict themselves to functions of one spatial variable, whose associated Wigner functions  are
easier to produce \cite{Rom} and draw \cite{Alo,Tes}. In the
present paper we will manage with Wigner
distributions in four variables, so that the effect of
diffraction will be a 4--rotation. Doing so, we will notice that the
4--rotation cannot be an arbitrary rotation: it is such that it can
split into two 2--rotations (as seen by Lohmann \cite{Loh}).

The (one-dimensional) fractional Fourier transformation of order $\alpha$
is sometimes defined as a rotation of angle $-\alpha$ in  phase space, generally the time--frequency space \cite{Alm}. 
Speaking of rotation in such a space makes no sense, indeed, as far
as appropriate scaled variables are not used. 

In the time--frequency
space, a rotation of angle $-\alpha$ would transform the $t$--$\nu$ axes into
$u$--$v$ axes, say, and if we use matrix notation, we should write
someting like
\begin{equation}
\begin{pmatrix}{u\cr v\cr}\end{pmatrix}=\begin{pmatrix}{\cos\alpha &  \sin\alpha\cr -\sin\alpha &  \cos\alpha}\end{pmatrix}\begin{pmatrix}{t\cr \nu}\end{pmatrix}\,,\end{equation}
and then $u=t\cos\alpha +\nu \sin\alpha$,
which makes no sense since $t$ is a time and $\nu$ a frequency, unless
appropriate scaled dimensionless  (or homogeneous) variables have been defined \cite{Oza3,Loh}.


We conclude that the effect of diffraction on the Wigner
distribution can be properly described only after choosing
appropriate scaling factors and thus, by defining an appropriate phase-space on
which rotations make sense. We propose a solution in
Sects. \ref{sect2}--\ref{sect5}
 and present a synthetic approach,
linking together diffraction phenomena, fractional order Fourier transformations and Wigner distributions.

Finally, we apply our theory to optical resonators and Gaussian beams. The main result
is that the Wigner distribution associated with the optical field inside a
resonator must be invariant by a 4--rotation of a particular kind. The whole theory of
stable optical resonators can be deduced from that one property: we deduce the
existence of transverse modes, represented by Hermite-Gauss
functions; we prove the existence of the waist and  provide usual
related formulae; we also  introduce the Gouy phase.

\section{Field transfer by 
  diffraction: real-order transfer}\label{sect2}
We use and adapt the representation of a 
diffraction
phenomenon by a fractional order Fourier transformation as developed in
various papers \cite{PPF1,PPF3,PPF5}, in the framework of a scalar theory.

\begin{figure}[h]
   \begin{center}
     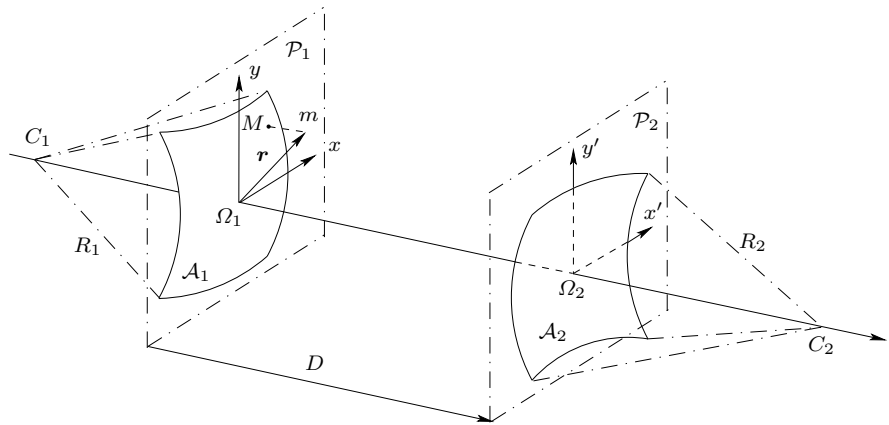
     \caption{Diffraction from a spherical  emitter  ${\cal A}_1$ to a spherical receiver ${\cal A}_2$ at a distance $D$. The point $M$ is represented by the coordinates $(x,y)$ of its projection $m$ on the plane ${\cal P}_1$ tangent to ${\cal A}_1$ at its vertex $\Omega_1$. \label{fig0} }
     \end{center}
 \end{figure}

We consider a spherical emitter ${\cal A}_1$ (Fig. \ref{fig0}), that is, a spherical
cap emitting  monochromatic light (wavelength 
$\lambda$ in the considered homogeneous and isotropic propagation medium):
if its vertex is $\Omega_1$
and its center of curvature $C_1$, the radius of curvature of ${\cal
  A}_1$ is $R_1=\overline{\Omega_1C_1}$,
 where $\overline{\Omega_1C_1}$ is an
algebraic measure. Algebraic measures are positive if taken in the
sense of light propagation. More generally, ${\cal A}_1$ can be
 an immaterial spherical cap, illuminated by a light wave.  We also consider a spherical receiver
${\cal A}_2$ (radius of curvature $R_2$) at a distance $D$ from ${\cal A}_1$ (we
write ``distance'', but $D$ is
 an algebraic measure ---we also use ``algebraic distance'': $D=\overline{\Omega_1\Omega_2}$, where
  $\Omega_2$ is the vertex of ${\cal A}_2$; distance $D$ is negative if ${\cal
   A}_2$ is a virtual receiver).
A point $M$ on ${\cal A}_1$ is represented by the coordinates $(x,y)$ of its projection $m$ on the plane tangent to ${\cal A}_1$ at its vertex $\Omega_1$ (Fig. \ref{fig0}).

Let $U_1$ denote the electric field amplitude on ${\cal A}_1$, and
$U_2$ the amplitude on
${\cal A}_2$. With Cartesian coordinates and vectorial notations $\vec r=(x,y)$ on ${\cal A}_1$ and $\vec r'=(x',y')$ on ${\cal A}_2$, the field transfer from ${\cal A}_1$ to ${\cal A}_2$ is expressed by\cite{PPF5}
\begin{eqnarray}
U_2(\vec r')\!\!\!\!&=&\!\!\!\!{\I\over \lambda D} \exp\!\!\left[-{\I\pi\over
 \lambda}\left({1\over R_2}+{1\over D}\right)\!r'^2\!\right] 
\!\int_{{\mathbb R}^2}\!\!\!\!\exp\!\left[-{\I\pi\over \lambda}\left({1\over D}-{1\over R_1}\right)\!r^2\!\right] 
\exp\!\left({2\I\pi\over \lambda D}\,\vec r\vec\cdot\vec
r'\right)\!U_1(\vec r)\, \D\vec r\,,\nonumber \\
& &\label{eq1}
\end{eqnarray}
where $\D \vec r=\D x\,\D y$ and $r=(x^2+y^2)^{1/2}$, and where $\vec r\vec\cdot\vec r'$ denotes the Euclidean scalar product of $\vec r$ and $\vec r'$.
A phase factor $\exp\, (-2\I\pi D/\lambda )$ has been omitted in Eq. (\ref{eq1}) (this factor
will be reintroduced  later on, when necessary).

We now write Eq. (\ref{eq1}) by using a fractional order Fourier
transformation. For a function $f$ of two variables, such a transformation is
defined by \cite{Nam}
\begin{equation}
{\cal F}_\alpha[f](\vec \rho ')={\I\E^{-\I\alpha}\over
  \sin\alpha}\,\exp\, (-\I\pi \rho'^2\cot\alpha) 
\int_{{\mathbb R^2}} \exp\, (-\I\pi \rho^2\cot\alpha) 
\,\exp\left({2\I\pi \vec \rho\vec \cdot\vec\rho '\over
  \sin\alpha}\right)\,f(\vec \rho )\,\D\vec\rho\,,\label{eq2}
  \end{equation}
where $\alpha$ is the order of the transformation, and $\vec \rho$ and
$\vec \rho '$ are 2--dimensional  real vectors without physical
dimension. The standard Fourier transformation is ${\cal F}_{\pi /2}$.

In the first part of the paper we will restrict ourselves to real
values of $\alpha$. The second part of the paper is devoted to complex
$\alpha$.

We define
\begin{equation}
J={(R_1-D)(D+R_2)\over D(D-R_1+R_2)}\,,\end{equation}   
and we assume $J\ge 0$ ($J<0$ is analyzed in the second part of the paper).

Then we define
the order $\alpha$ of the transformation associated with diffraction from
${\cal A}_1$ to ${\cal A}_2$, as expressed by Eq. (\ref{eq1}),
by 
\begin{equation}
\cot^2\alpha=J\,,\label{eq3}\end{equation}
with $-\pi < \alpha <\pi$, and $\alpha D\ge 0$.

We remark that Eq. (\ref{eq3}) is equivalent to
\begin{equation} \cos^2\alpha =\left(1-{D\over
   R_1}\right)\left(1+{D\over R_2}\right)\,,
\end{equation}
which can also be used in defining $\alpha$.

So far, the sign of $\cot\alpha$ is not determined. 
We then introduce the auxiliary parameter $\varepsilon_1$  such that
\begin{equation}
\varepsilon_1={D\over R_1-D}\cot\alpha\,,\hskip 1cm
\varepsilon_1 R_1>0 \,,\label{equ2}\end{equation} 
and which determines the sign of $\cot\alpha$, so that $\alpha$ is also
totally determined, after taking into account the previous definitions.

At last, we define the auxiliary parameter $\varepsilon_2$ such that 
\begin{equation}
\varepsilon_2={D\over R_2+D}\cot\alpha\,.\label{equ2b}\end{equation} 
In Appendix A, we show that $\varepsilon_2 R_2>0$.

One of the main points in associating a fractional order Fourier transformation
with a diffraction phenomenon is the choice of sca\-led vectorial
variables on the emitter and on the receiver. We choose
\begin{equation}
\vec \rho ={\vec r\over \sqrt{\lambda \varepsilon_1 R_1}}\,, \hskip
.5cm \mbox{and}\hskip .5cm 
\vec \rho '={\vec r'\over \sqrt{\lambda \varepsilon_2 R_2}}\,,\end{equation}
on ${\cal A}_1$ and ${\cal A}_2$ respectively. We remark that both $\vec \rho$ and $\vec \rho '$ are
physically dimensionless.

We also use the following scaled field amplitudes on ${\cal A}_1$ and ${\cal A}_2$
\begin{equation}
V_1(\vec \rho )=\sqrt{\varepsilon_1R_1\over
  \lambda}\;U_1\left(\sqrt{\lambda \varepsilon_1 R_1}\,\vec
\rho\right)\,,\label{eq10}\end{equation}
\begin{equation}
V_2(\vec \rho ')=\sqrt{\varepsilon_2R_2\over \lambda}\;U_2\left(\sqrt{\lambda \varepsilon_2 R_2}\,\vec \rho '\right)\,.\label{eq11}\end{equation}
Then Eq. (\ref{eq1}) becomes
\begin{equation}
V_2=\E^{\I\alpha}{\cal F}_\alpha [V_1]\,,\label{eq7}\end{equation}
which expresses the field transfer from ${\cal A}_1$ to ${\cal A}_2$
through a fractional order Fourier transformation. 

Generally Eq. $\!$(\ref{eq7}) corresponds to a Fresnel
phenomenon. Fraunhofer diffraction is a limit case, obtained for $\alpha =\pi /2$.

\section{Transfer of the spherical angular spectrum}\label{sect3}

The spherical angular spectrum $S$ of the field amplitude $U$ on the
spherical emitter (or receiver) ${\cal A}$ is defined by \cite{PPF4,PPF4b}
\begin{equation}
S(\vec \Phi )={1\over \lambda^2}\;\widehat U\!\left({\vec \Phi \over \lambda}\right)\,,
\end{equation}
where $\widehat U$ denotes the Fourier transform of $U$,
and the 2--dimensional vectorial variable $\vec \Phi$ is the angular spatial frequen\-cy, related to the spatial frequency $\vec F$ by $\vec\Phi =\lambda \vec F$.

The propagation of the spherical angular spectrum from the emitter
${\cal A}_1$ to the receiver ${\cal A}_2$ at a distance $D$  is given by \cite{PPF4,PPF4b}
\begin{eqnarray}
S_2(\vec \Phi ')\!\!\!\!&=&\!\!\!\!{\I R_1R_2 \over \lambda (D-R_1+R_2)}\exp\left(-{\I\pi R_2 (R_1-D)\over \lambda (D-R_1+R_2)}\,\Phi'^2\right) \label{eq14} \\
&&\times \;\int_{{\mathbb R}^2}\exp\left(-{\I\pi R_1(D+R_2)\over \lambda (D-R_1+R_2)}\,\Phi^2\right) \, \exp\left({2\I\pi  R_1R_2 \over\lambda(D-R_1+R_2)}\,\vec \Phi \vec \cdot\vec \Phi '\right) 
S_1(\vec \Phi )\,\D \vec \Phi \,.\nonumber
\end{eqnarray}

We remark that Eq. (\ref{eq14}) is similar to Eq. (\ref{eq1}), where
$\vec r$ and $\vec r'$ are replaced by $\vec \Phi$ and $\vec \Phi '$
and where $D$, $R_1$ and $R_2$ are changed according to
\begin{equation}
D\longmapsto {D-R_1+R_2\over R_1R_2}\,,\label{eq15}\end{equation}
\begin{equation}
R_1\longmapsto -{D-R_1+R_2\over R_1D}\,,\label{eq16}\end{equation}
\begin{equation}
R_2\longmapsto -{D-R_1+R_2\over R_2D}\,.\label{eq17}\end{equation}

We  choose $\alpha$,
$\varepsilon_1$ and $\varepsilon_2$ as in Eqs. (\ref{eq3}),
(\ref{equ2}) and (\ref{equ2b}).  By introducing the following scaled angular
variables on ${\cal A}_1$ and on ${\cal A}_2$ respectively
\begin{equation}
 \vec \phi =\sqrt{\varepsilon_1 R_1\over \lambda}\,\vec \Phi\,,\hskip
.5cm \mbox{and}\hskip .5cm
\vec \phi '=\sqrt{\varepsilon_2 R_2\over \lambda}\,\vec \Phi '\,,\end{equation}
and the scaled spherical angular spectra
\begin{equation}
T_1 (\vec \phi )=\sqrt{\lambda\over \varepsilon_1R_1}\,S_1\left(\sqrt{\lambda\over \varepsilon_1 R_1}\,\vec \phi\right)\,,\hskip
.5cm \mbox{and}\hskip .5cm 
T_2 (\vec \phi ')=\sqrt{\lambda\over \varepsilon_2R_2}\,S_2\left(\sqrt{\lambda\over \varepsilon_2 R_2}\,\vec \phi '\right)\,,\end{equation}
it can be proved \cite{PPF4,PPF4b} that Eq. (\ref{eq14}) becomes
\begin{equation}
T_2=\E^{\I\alpha}{\cal F}_\alpha [T_1]\,.\label{eq23}\end{equation}

Eq. (\ref{eq23})  is identical to Eq. (\ref{eq7}): the same fractional
order Fourier transformation expresses the scaled
spherical angular spectrum propagation as well as the scaled field
amplitude propagation.

Both $\vec \phi$ and $\vec \phi '$ are 2--dimensional
vectors without physical dimensions. We  have
\begin{equation}
 \vec r\vec\cdot \vec F={1\over  \lambda} \, \vec r \vec\cdot\vec\Phi
 =\vec \rho\vec
 \cdot\vec\phi\,,\label{eq21}\end{equation}
\begin{equation}
 \vec r'\vec\cdot \vec F'={1\over  \lambda} \,\vec r' \vec\cdot\vec\Phi '  =\,\vec \rho '\vec \cdot\vec\phi '\,,\label{eq22}\end{equation}
so that $\vec \phi$ (resp. $\vec \phi '$) is the conjugate variable of
$\vec \rho$ (resp. $\vec\rho '$). Eqs. (\ref{eq21}) and (\ref{eq22})
make sense if rational units are used both for lengths and spatial
frequencies (for example
mm and mm$^{-1}$).

We point out that $T_1$ is no more than the (2--dimen\-sional) Fourier transform of $V_1$: from Eq. (\ref{eq10}), indeed,  we deduce
\begin{equation}
\widehat V_1(\vec \phi )={1\over  \lambda\sqrt{\lambda\varepsilon_1R_1}}\,\widehat U_1\left({\vec \phi \over \sqrt{\lambda \varepsilon_1R_1}}\right)
=\sqrt{\lambda \over \varepsilon_1R_1}\,S_1\left(\sqrt{\lambda\over \varepsilon_1R_1}\vec \phi\right) =T_1(\vec \phi )\,.
\end{equation}
Of course $T_2$ is the Fourier transform of $V_2$, and finally, Eq. (\ref{eq23}) can  be written
\begin{equation}
\widehat V_2=\E^{\I\alpha}{\cal F}_\alpha \bigl[\widehat V_1\bigr]\,.\label{eq23b}\end{equation}
Eq. (\ref{eq23b}) can also be deduced from Eq. (\ref{eq7}) by using
the commutativity of the product of fractional Fourier transformations  (${\cal F}_\alpha \circ
{\cal F}_\beta = {\cal F}_\beta \circ
{\cal F}_\alpha$).

The propagation of the optical field is summarized in the following diagram
\begin{equation}
  \hskip .63cm
\begin{CD}
 V_1 @> \E^{\I\alpha}{\cal F}_\alpha >> V_2\\
 @V {\cal F}_{\pi /2} VV @VV {\cal F}_{\pi /2} V  \\
 \widehat V_1@> \E^{\I\alpha}{\cal F}_\alpha >>  \widehat V_2\\
\end{CD}
\label{eq6n}
\end{equation}
where the symmetry between propagations of the field amplitude and of
the spherical angular spectrum is conspicuous.

\section{Wigner distribution associated with the field}\label{sect4}

The Wigner distribution of a $x$--function $f$,  is a function $W(x,y)$, where $y$ is the conjugate variable of $x$, and $W(x,y)$ represents the localization of $f$ in the phase space $x$--$y$.

To obtain the Wigner
distribution corresponding to an optical field---whose amplitude is referred to spatial variables---it is 
natural to use also the angular spectrum, since it represents   the field in the domain of spatial  frequencies, which are the conjugates of spatial
variables. 


There is, indeed, a basic reason for using the spherical angular spectrum
instead of the planar angular spectrum as usually defined in Fourier
Optics \cite{Goo}: the spherical angular spectrum, unlike the planar
angular spectrum, propagates in the same way as the field amplitude,
as shown in Sect. \ref{sect3}, Eqs. (\ref{eq7}) and (\ref{eq23}) and
diagram (\ref{eq6n}).  In Fourier optics \cite{Goo}, and considering
diffraction between two planes, propagation of the optical
field generally corresponds to the following diagram 
\begin{equation}
   \hskip .63cm
\begin{CD}
 U_1 @> *h >> U_2\\
 @V {\cal F}_{\pi /2} VV @VV {\cal F}_{\pi /2} V  \\
 \widehat U_1@> \times H >>  \widehat U_2\\
\end{CD}
\label{eq6nb}
\end{equation}
where $\times H$ denotes the multiplication by a transfer function of
the form $H(\vec F)=\exp (\I\pi \lambda DF^2)$, and $*h$ denotes the
convolution product by 
$h(\vec r)=(\I /\lambda D) \exp (-\I\pi r^2/\lambda D)$. The symmetry of diagram
(\ref{eq6n}) is broken.

The previous difference between the two angular spectra
is an important feature to be taken into account in defining the
Wigner representation.  The effect of diffraction in the phase space will be homogeneous
 only if the Wigner distribution is related to 
the spherical angular spectrum.

We then define the Wigner distribution of the optical field on a
spherical cap ${\cal A}$ as the Wigner distribution of the scaled
field amplitude $V$ on ${\cal A}$, that is, 
\begin{equation}
W(\vec \rho ,\vec \phi )=\int_{{\mathbb R}^2}\!\!V\left(\vec \rho +{\vec \tau \over 2}\right)
\overline{V\left(\vec \rho -{\vec \tau \over 2}\right)}
\,\E^{2\I\pi\vec \tau\vec\cdot\vec\phi} \,\D\vec\tau\,,\label{eq27}\end{equation}
where $\overline{V}$ denotes the complex conjugate of $V$.

The Wigner distribution is defined in the 4--dimensional ph\-a\-se space
$\vec \rho$--$\vec \phi$, which has no physical dimension and which
will be called the ``scaled phase-space''.

It can be proved that
\begin{equation}
W(\vec \rho ,\vec \phi )=\int_{{\mathbb R}^2}\!\!\widehat V\left(\vec \phi +{\vec \eta \over 2}\right)
\overline{\widehat V\left(\vec \phi -{\vec \eta \over 2}\right)}
\,
\E^{-2\I\pi  \vec \rho \vec\cdot\vec\eta}\,\D\vec\eta\,,\label{eq28}\end{equation}
so that  the above defined Wigner distribution  has the usual properties
assigned to Wigner distributions.  We also have
\begin{equation}
\int_{{\mathbb R}^2}\!\!W(\vec \rho ,\vec \phi )\,\D\vec\phi=|V(\vec \rho
)|^2\,,
\hskip .5cm {\rm and}\hskip .5cm 
\int_{{\mathbb R}^2}\!\!W(\vec \rho ,\vec \phi )\,\D\vec\rho = |\widehat V(\vec \phi )|^2\,.\label{eq28n}\end{equation}

\section{Effect of 
  diffraction on the Wigner distribution:  real-order transfer}\label{sect5}

Let $\alpha$ be the order of the fractional
Fourier transformation associated with the field transfer from the
spherical emitter ${\cal A}_1$ to the spherical receiver ${\cal
  A}_2$ (see Sect. \ref{sect2}). Let $W_j$ denote
the Wigner distribution associated with the field amplitude $U_j$ on ${\cal A}_j$ ($j=1,2$). The effect of diffraction on the Wigner distribution is expressed by
\begin{equation}
W_2(\vec\rho ,\vec \phi )=W_1(\vec \rho \cos\alpha -\vec
\phi\sin\alpha ,\vec\rho\sin\alpha +\vec\phi\cos\alpha
)\,.\label{eq25}\end{equation}

A proof is as follows. For sake of conciseness we define 
$E(x)=\exp (\I\pi x)$.
We use Eqs. (\ref{eq2}) and (\ref{eq7})  and write
\begin{eqnarray}
W_2(\vec \rho, \vec \phi )\!\!\!\!&=&\!\!\!\!{1\over \sin^2\alpha}
\int_{{\mathbb R}^2} E \left(-\left\Vert \vec \rho +{\vec \tau  \over 2}\right\Vert^2\cot\alpha\right)\nonumber\\
&&\times  \int_{{\mathbb R}^2}  E(-\rho
'^2\cot\alpha)\,E\!\left[{2\vec\rho
    '\over\sin\alpha}\!\vec\cdot\!\left(\vec\rho+{\vec\tau\over
    2}\right)\right]  \;E \left(\left\Vert \vec \rho -{\vec
  \tau\over 2}\right\Vert^2\cot\alpha\right) V_1(\vec\rho ')\,\D\vec\rho '
\nonumber \\
&&\times
 \int_{{\mathbb R}^2} E\left[-{2\vec\rho ''\over\sin\alpha}\vec\cdot\left(\vec\rho-{\vec\tau\over 2}\right)\right] E(\rho ''^2\cot\alpha)\,\overline{V_1(\vec \rho '')}\,\D \vec \rho '' \,E(2\vec\tau\vec\cdot\vec\phi  )\,\D\vec \tau \nonumber \\
&=&\!\!\!\!{1\over \sin^2\alpha}
\int_{{\mathbb R}^2} E(-\rho'^2\cot\alpha )\,E\!\left({2\vec \rho\vec\cdot\vec\rho '\over \sin\alpha}\right)\,V_1(\vec\rho ')\,\D\vec\rho '\nonumber \\
&&\times \;\int_{{\mathbb R}^2}\!\!\!E(\rho''^2\cot\alpha )\,E\!\left(-{2\vec \rho\vec\cdot\vec\rho ''\over \sin\alpha}\right)\,\overline{V_1(\vec\rho '')}\,\D\vec\rho ''\nonumber \\
&&\times \;\int_{{\mathbb R}^2}\!\!E(-2\vec\rho\vec\cdot\vec\tau\cot\alpha )
E\!\left({\vec\rho '+\vec\rho ''\over\sin\alpha}\vec\cdot\vec\tau\right)
\,E(2\vec\tau\vec\cdot\vec\phi )\,\D\vec\tau\,.
\label{eq32}
\end{eqnarray}
If $\delta$ denotes the Dirac generalized function, the last integral in Eq. (\ref{eq32}) is equal to
\begin{equation}
\delta \left(\vec\phi -\vec\rho\cot\alpha +{\vec\rho '+\vec\rho ''\over 2\sin\alpha}\right)
 =4\sin^2\alpha\,\delta \bigl(2\vec\phi\sin\alpha -2\vec\rho\cos\alpha +\vec\rho '+\vec\rho ''\bigr)\,,
\end{equation}
so that Eq. (\ref{eq32}) becomes
\begin{eqnarray}
W_2(\vec \rho, \vec \phi )\!\!\!\!&=&\!\!\!\!
4\int_{{\mathbb R}^2}E(-\rho'^2\cot\alpha )\,E\!\left({2\vec\rho\vec\cdot\vec\rho '\over\sin\alpha}\right) E\bigl(\Vert 2\vec\rho\cos\alpha -2\vec\phi\sin\alpha -\vec\rho '\Vert^2\cot\alpha \bigr)
\label{eq33} \\
&&\hskip -.3cm \times \;E\left[-{2\vec \rho\over\sin\alpha}\vec\cdot \bigl(2\vec\rho\cos\alpha -2\vec\phi\sin\alpha -\vec\rho '\bigr)\right]
V_1(\vec\rho ') \,\overline{V_1(2\vec\rho\cos\alpha -2\vec \phi\sin\alpha -\vec\rho ')}\,\D\vec\rho '\,.
\nonumber
\end{eqnarray}
We change $\vec\rho '$ into
$\vec \tau =2\vec \rho '-2\vec\rho \cos\alpha+2\vec\phi\sin\alpha$,
so that Eq. (\ref{eq33}) becomes
\begin{eqnarray}
W_2(\vec \rho, \vec \phi )\!\!\!\! &=&\!\!\!\!\int_{{\mathbb R}^2}
V_1\left(\vec\rho \cos\alpha -\vec\phi\sin\alpha +{\vec\tau\over 2}\right)
 \overline{V_1\left(\vec\rho \cos\alpha -\vec\phi\sin\alpha
   -{\vec\tau\over 2}\right)}\nonumber  \\
& & \hskip 6cm \times\;
 E\bigl[2(\vec\rho\sin\alpha +\vec\phi \cos\alpha )\vec\cdot\vec\tau\bigr] \,\D\vec \tau\,,
\end{eqnarray}
which is eq. (\ref{eq25}) once more. The proof is complete.

Equation (\ref{eq25}) shows that the effect of a 
diffraction
phenomenon is a rotation of the Wigner distribution associated with the
scaled field amplitude. The rotation operates in the 4--dimensional scaled
phase-space $\vec \rho$--$\vec \phi$. The value of the function $W_2$ at point
$(\vec \rho , \vec{\phi})$ is equal to the value of the function $W_1$ at point 
$(\vec \rho \cos\alpha -\vec \phi \sin\alpha , \vec \rho \sin\alpha
+\vec\phi\cos\alpha )$, so  that the ``angle'' of the rotation is
equal to $-\alpha$, and is opposite to the order of the fractional Fourier
transformation associated with the field amplitude transfer (see Appendix B).

We remark that the 4--dimensional scaled phase-space is homogeneous to ${\mathbb
  R}^4$. It is equipped with an Euclidean norm defined by
\begin{equation}
||(\vec \rho ,\vec \phi)||^2=\rho_x^2+\rho_y^2+\phi_x^2+\phi_y^2\,,
\end{equation}
where $\vec \rho =(\rho_x,\rho_y)$ and $\vec \phi =(\phi_x,\phi_y)$.
Then the above mentioned rotation makes sense.

A matrix representation of the effect of diffraction can be seen as a
coordinate transformation in the scaled phase-space and is the following. We use $\vec \rho =(\rho_x,\rho_y)$ and $\vec \phi =(\phi_x,\phi_y)$ and a $1$ or $2$ index for the emitter ${\cal A}_1$ or the receiver ${\cal A}_2$. Then
\begin{equation}
\begin{pmatrix}{\rho_{x2}\cr \rho_{y2}\cr \phi_{x2}\cr \phi_{y2}}\end{pmatrix}=
\begin{pmatrix}{\cos\alpha & 0 &\sin\alpha & 0 \cr
0 & \cos\alpha  & 0 & \sin\alpha \cr
    -\sin\alpha & 0 &\cos\alpha &0 \cr
 0 &-\sin\alpha & 0 &\cos\alpha }\end{pmatrix}
\begin{pmatrix}{\rho_{x1}\cr \rho_{y1}\cr \phi_{x1}\cr
    \phi_{y1}}\end{pmatrix}\,.
\label{eq36a}
\end{equation}
An equivalent  matrix form is obtained by reordering the variables, that is,
\begin{equation}
\begin{pmatrix}{\rho_{x2}\cr \phi_{x2}\cr \rho_{y2}\cr \phi_{y2}}\end{pmatrix}=
\begin{pmatrix}{\cos\alpha & \sin\alpha & 0 &0 \cr -\sin\alpha &\cos\alpha &0 &0\cr
 0 &0 &\cos\alpha & \sin\alpha \cr 0 &0& -\sin\alpha &\cos\alpha }\end{pmatrix}
\begin{pmatrix}{\rho_{x1}\cr \phi_{x1}\cr \rho_{y1}\cr
    \phi_{y1}}\end{pmatrix}\,.
\label{eq36}
\end{equation}
In the following, we call ``Wigner rotation of parameter (or angle) $-\alpha$'' a 4--dimensional rotation
whose matrix is given by Eq. (\ref{eq36}). We denote it by ${\cal R}_{-\alpha}$.

The 4-rotation matrix can be written as the (commutative) product of two matrices
\begin{equation}
\begin{pmatrix}{\cos\alpha & \sin\alpha & 0 &0 \cr -\sin\alpha &\cos\alpha &0 &0\cr
 0 &0 &1 & 0 \cr 0 &0& 0 &1
 }\end{pmatrix}
\begin{pmatrix}{1 & 0 & 0 &0 \cr 0&1 &0 &0\cr
 0 &0 &\cos\alpha & \sin\alpha \cr 0 &0& -\sin\alpha &\cos\alpha }\end{pmatrix}\,,\label{eq37b}
\end{equation}
and each matrix represents a rotation on a 2--dimensional subspace of
${\mathbb R}^4$, that is, a plane. In these planes, the effect of
diffraction is expressed through a 2--rotation of angle $-\alpha$, whose matrix is
\begin{equation}
\begin{pmatrix}{\rho_{k2}\cr \phi_{k2}\cr}\end{pmatrix}=
\begin{pmatrix}{\cos\alpha & \sin\alpha\cr -\sin\alpha &\cos\alpha\cr }\end{pmatrix}
\begin{pmatrix}{\rho_{k1}\cr
    \phi_{k1}\cr}\end{pmatrix}\;\,\mbox{where}\; k=x,y.\nonumber \\
\label{eq37}
\end{equation} 
This will help in concretely representing the effect of
diffraction on the Wigner distribution.

We conclude that the matrix representation of the effect of 
dif\-frac\-tion on the Wigner distribution in 
the scaled phase-space is not an arbitrary rotation matrix, but a matrix that
can split according to (\ref{eq37b}).

\section{Composition of two transformations}\label{sect6}

According to the Huygens--Fresnel principle, the field transfer from an emitter
${\cal A}_1$ to a receiver ${\cal A}_2$ can be thought of as the composition of
two transfers: from ${\cal A}_1$ to ${\cal A}_3$ and from ${\cal A}_3$ to ${\cal A}_2$,
where ${\cal A}_3$ is an intermediate spherical cap located between ${\cal
  A}_1$ and ${\cal A}_2$. 

More generally, ${\cal A}_3$ can be every (immaterial) spherical cap, not
necessarily located between ${\cal A}_1$ and ${\cal A}_2$. For example
if light encounters first ${\cal A}_1$, then ${\cal A}_2$ and 
${\cal A}_3$ at last,  the
transfer from ${\cal A}_3$ to ${\cal A}_2$ is virtual.

We denote $D_1$ the (algebraic) distance from ${\cal A}_1$ to ${\cal A}_3$ and
$D_2$ from ${\cal A}_3$ to ${\cal A}_2$; the (algebraic) distance from ${\cal
  A}_1$ to ${\cal A}_2$ is $D=D_1+D_2$ (this relation holds true even
if ${\cal A}_3$ is not between ${\cal A}_1$ and ${\cal A}_2$; for
example $D_2$ is negative when the transfer from ${\cal A}_3$ to ${\cal
  A}_2$ is virtual).

The transfer from ${\cal A}_1$ to ${\cal A}_3$ is described by a
fractional Fourier transformation whose order is $\alpha_1$ and the
transfer from ${\cal A}_3$ to ${\cal A}_2$ by a transformation whose order
is $\alpha_2$. According to the Huygens--Fresnel principle the field
transfer from ${\cal A}_1$ to ${\cal A}_2$ is expressed by a fractional
Fourier transformation whose order is $\alpha =\alpha_1+\alpha_2$ \cite{Fog2}.
Interpreted in terms of Wigner representation, the
Huygens--Fresnel principle is equivalent to the composition of two
Wigner rotations whose respective parameters are $-\alpha_1$ and $-\alpha_2$, which
results in a Wigner rotation whose parameter is $-\alpha =-\alpha_1
-\alpha_2$. 

The previous composition makes senses only if scaled
variables on ${\cal A}_3$ associated with the first rotation are also scaled
variables for the second rotation. 
We introduce parameters $\varepsilon_{11}$ and $\varepsilon_{12}$
(transfer from ${\cal A}_1$ to ${\cal A}_3$) and
$\varepsilon_{21}$ and $\varepsilon_{22}$ (transfer from ${\cal A}_3$ to ${\cal A}_2$). Scaled variables on ${\cal A}_3$ must be the same
for both transfers, which implies $\varepsilon_{12}=\varepsilon_{21}$, that is,
\begin{equation}
{D_1(R_1-D_1)\over (R_3+D_1)(D_1-R_1+R_3)}={D_2(R_2+D_2)\over (R_3-D_2)
 (D_2-R_3+R_2)}.\label{equ37}\end{equation}
Eq. (\ref{equ37}) reduces, indeed, to a first degree equation in $R_3$, whose
  solution is \cite{Fog2}
\begin{equation}
R_3={D_1(R_2+D_2)(R_1-D)+D_2(R_2+D)(R_1-D_1)\over
  D_1(R_1-D)+D_2(R_2+D)}\,.\label{equ38}\end{equation}

Finally, if $V_j$ ($j=1,2,3$) denotes the scaled field amplitude on ${\cal A}_j$,  the Huygens--Fresnel principle is expressed by
\begin{equation}
V_2=\E^{\I\alpha_2}{\cal F}_{\alpha_2}[V_3]=\E^{\I\alpha_2}\E^{\I\alpha_1}{\cal
  F}_{\alpha_2}\circ{\cal F}_{\alpha_1} [V_1]=\E^{\I\alpha}{\cal
  F}_\alpha[V_1]\,,
\end{equation}
where  $\alpha =\alpha_1+\alpha_2$.
Since fractional order Fourier transformations commute, we also have
\begin{equation}
\widehat V_2={\cal F}_{\pi /2} [V_2]=\E^{\I\alpha_2}{\cal
  F}_{\alpha_2}\circ{\cal F}_{\pi /2}[V_3]=\E^{\I\alpha_2}{\cal
  F}_{\alpha_2}[\widehat V_3]=\E^{\I\alpha_2}\E^{\I\alpha_1}{\cal
  F}_{\alpha_2}\circ{\cal F}_{\alpha_1} [\widehat V_1]=\E^{\I\alpha}{\cal
  F}_\alpha[\widehat V_1]\,.
\end{equation}

If $W_j$ denotes the  Wigner distribution on ${\cal A}_j$, we draw
the following diagrams
\begin{equation}\bfig
\Vtriangle/>`>`<-/<400,400>[V_1`V_2`V_3;\E^{\I\alpha}{\cal F}_\alpha `\E^{\I\alpha_1}{\cal
    F}_{\alpha_1}`\E^{\I\alpha_2}{\cal F}_{\alpha_2}]
\hskip 4.5cm
\Vtriangle/>`>`<-/<400,400>[\widehat V_1`\widehat V_2`\widehat V_3;\E^{\I\alpha}{\cal F}_\alpha `\E^{\I\alpha_1}{\cal
    F}_{\alpha_1}`\E^{\I\alpha_2}{\cal F}_{\alpha_2}]
\hskip 4.5cm
\Vtriangle/>`>`<-/<400,400>[W_1`W_2`W_3;{\cal R}_{-\alpha} `{\cal R}_{-\alpha_1}`{\cal R}_{-\alpha_2}]
\efig
\end{equation}
which express the Huygens-Fresnel principle.
For every spherical cap (${\cal A}_1$, ${\cal A}_2$ or ${\cal
  A}_3$) two fractional Fourier
transformations are involved that apply to a same scaled amplitude (and
then the same scaled variable). For example $V_1$ is the input scaled
amplitude on ${\cal A}_1$ for both ${\cal F}_\alpha$ and  ${\cal F}_{\alpha_1}$.

\section{Application to optical resonator theory}\label{sect7}
\subsection{Sign conventions}
 Since we are going to consider  mirrors, we will adopt the following sign
 conventions. We maintain the rule that an algebraic measure is
 positive if taken in the sense of light propagation. If a mirror has
 a vertex $\Omega$ and a center of curvature $C$, its radius is
 $R=\overline{\Omega C}$. For a concave mirror, $\overline {\Omega C}$
 is negative if we consider light propagation before reflection on the
 mirror (since light travels to the mirror, that is from $C$ to
 $\Omega$); it is positive after reflection (since light goes away from
 the mirror, that is, from $\Omega$ to $C$). For that reason we associate
 two radii with a mirror: the objet radius, denoted $R$, is related to
 light propagation before reflection on the mirror; the image radius, denoted $R'$, is
 related to light propagation after reflection. We have $R'=-R$.

 \subsection{Round trip}\label{sect72}
We consider an optical resonator made up of two spherical mirrors
${\cal M}_1$ and ${\cal M}_2$. The object radius of ${\cal M}_1$ is
$R_1$ and its image radius is $R'_1$. They are $R_2$ and $R'_2$ for
${\cal M}_2$. 

For diffraction from ${\cal M}_1$ to ${\cal M}_2$, the
distance to be considered is the distance from ${\cal M}_1$ to ${\cal
  M}_2$, say $D$, after reflection on ${\cal M}_1$. For
diffraction from ${\cal M}_2$ to ${\cal M}_1$, the distance to be
considered, say $D'$, is taken
from ${\cal M}_2$ to ${\cal M}_1$ and is related to light
propagation after reflection on ${\cal M}_2$. According to our sign
convention, we have $D=D'$. We then introduce the ``length'' of the resonator,
which is $L=D=D'$, and which is independent of the sense of ligth
propagation ($L$ can be negative for a virtual resonator ;  $L$
should be called ``algebraic length'').

We consider first the field
transfer from ${\cal M}_1$ to ${\cal M}_2$ and will apply the result
of Sect. \ref{sect2}. Light propagates from ${\cal M}_1$ to
${\cal M}_2$ after reflection on ${\cal M}_1$ and before reflection on
${\cal M}_2$: according to our convention, the radii to be considered
are $R'_1$ and $R_2$. Then the field transfer from ${\cal M}_1$ to
${\cal M}_2$
is described with the help of
$\alpha_0$ such that
\begin{equation}
\cot^2\alpha_0={(R'_1-L)(L+R_2)\over L(L-R'_1+R_2)}\,,\label{eq39}\end{equation}
with $\alpha_0 L\ge 0$ (we assume that $\alpha_0$ is a real number). We also have
\begin{equation}
\varepsilon_1={L\over R'_1-L}\cot\alpha_0\,,\hskip 1cm \varepsilon_1 R'_1>0 \,,\label{equ40}\end{equation} and 
\begin{equation}
\varepsilon_2={L\over R_2+L}\cot\alpha_0\,,\hskip 1cm
 \varepsilon_2 R_2>0 \,.\label{equ40b}\end{equation}
With appropriate scaled variables on ${\cal M}_1$ and ${\cal M}_2$ we have
\begin{equation}
V_2=\E^{\I\alpha_0}{\cal F}_{\alpha_0} [V_1]\,.\label{equ41}\end{equation}

For the field transfer from ${\cal M}_2$ to ${\cal M}_1$, the emitter
is ${\cal M}_2$ and the receiver is ${\cal M}_1$; light propagates
after reflection on ${\cal M}_2$ and before reflection on ${\cal
  M}_1$. The radii to be considered are $R'_2$ for the emitter and
$R_1$ for the receiver. Then the  field transfer from ${\cal M}_2$ to ${\cal M}_1$ is described with
the help of $\alpha'_0$ such that
\begin{equation}
\cot^2\alpha '_0={(R'_2-L)(L+R_1)\over
  L(L-R'_2+R_1)}\,,\hskip .7cm \alpha_0'L\ge 0\,,\label{eq41}\end{equation}
\begin{equation}
\varepsilon'_1={L\over R'_2-L}\cot\alpha'_0\,,\hskip 1cm \varepsilon'_1 R'_2>0 \,,\label{equ42}\end{equation} and 
\begin{equation}
\varepsilon'_2={L\over R_1+L}\cot\alpha'_0\,,\hskip 1cm
 \varepsilon'_2 R_1>0 \,.\label{equ42b}\end{equation}

Since $R'_1=-R_1$ and $R'_2=-R_2$, then $\alpha_0 =\alpha '_0$. We
 also have
 $\varepsilon_1R'_1=\varepsilon'_2R_1$ and
 $\varepsilon_2R_2=\varepsilon'_1R'_2$, so that scaled variables on
 ${\cal M}_1$ and ${\cal M}_2$ are identical for both transfers:
 their composition makes sense. Finally,
 the transfer from ${\cal M}_1$ to ${\cal M}_1$ in a round trip (that
 is, after a reflection on ${\cal M}_2$) is
 expressed by
\begin{equation}
V_1=\E^{2\I\alpha_0}{\cal F}_{2\alpha_0} [V_1]\,,\label{eq43}\end{equation}
and the transfer from ${\cal M}_2$ to ${\cal M}_2$
by \begin{equation}
V_2=\E^{2\I\alpha_0}{\cal F}_{2\alpha_0}
[V_2]\,.\label{eq43bis}\end{equation}

\subsection{Eigenmodes} \label{sect73}
We consider an intermediate spherical cap ${\cal S}_\alpha$ at a distance
$D_{1}$ from ${\cal M}_1$. The distance from ${\cal S}_\alpha$
to ${\cal M}_2$ is $D_{2}$ and the length of the resonator is such that $L=D_{1}+D_{2}$.  Let $\alpha_1$ be the order of the fractional Fourier transformation
associated with the field transfer from ${\cal M}_1$ to ${\cal
  S}_\alpha$.

We choose the radius $R_\alpha$ of
${\cal S}_\alpha$ such that composition of transfers from ${\cal
  M}_1$ to ${\cal S}_\alpha$ and from ${\cal S}_\alpha$ to ${\cal
    M}_2$ makes sense, that is, according to Eq. (\ref{equ38})
\begin{equation}
R_\alpha ={D_{1}(R_2+D_{2})(R'_1-L)+D_{2}(R_2+L)(R'_1-D_{1})\over
  D_{1}(R'_1-L)+D_{2}(R_2+L)}\,.
\label{equ38bis}\end{equation}
Then scaled variables on ${\cal M}_1$ are identical for both transfers
from ${\cal M}_1$ to ${\cal S}_\alpha$ and from ${\cal
  M}_1$ to ${\cal M}_2$, so that the scaled amplitude on ${\cal
  S}_\alpha$ is $V_{\alpha_1}$ such that
\begin{equation}
V_{\alpha_1}=\E^{\I\alpha_1}{\cal F}_{\alpha_1}
[V_1]\,,\label{eq53n}\end{equation}
where $V_1$ is the $V_1$ of Eq. (\ref{eq43}).

We imagine now that ${\cal S}_\alpha$ is a mirror, without changing
its radius. A round trip from
${\cal M}_1$ to ${\cal M}_1$ after reflection on ${\cal S}_\alpha$ is
expressed by 
\begin{equation}
V_1=\E^{2\I\alpha_1}{\cal F}_ {2\alpha_1}
[V_1]\,.\label{equ49}\end{equation}

Since ${\cal S}_\alpha$ is an arbitrary spherical cap (not
nece\-ssa\-ri\-ly located between ${\cal M}_1$ and ${\cal M}_2$), with appropriate radius, the order
$\alpha_1$ can be an arbitrary real number.
We conclude that the scaled field amplitude $V_1$ on ${\cal M}_1$ is invariant in every
fractional order Fourier transformation, whatever its order. Since
Hermite--Gauss functions are eigenfunctions of all fractional order
Fou\-rier trans\-formations \cite{Nam}, then $V_1$ is
represented by a Hermite--Gauss function, or by a linear combination of
such functions (explicit expressions of
Hermite--Gauss functions will be given in Sect. \ref{sect92}).

 According to Eq. (\ref{eq53n}), the field amplitude on ${\cal
  S}_\alpha$ is obtained from the amplitude on ${\cal M}_1$ by
applying a fractional order Fourier transformation, so that the
scaled field amplitude on ${\cal S}_\alpha$ is expressed by the same Hermite--Gauss
function as the amplitude on ${\cal M}_1$. Since ${\cal S}_\alpha$ is
an arbitrary spherical cap (with appropriate radius), we conclude that for
every spherical surface of the family $\{{\cal S}_\alpha \}$ the
  amplitude is expressed by a Hermite--Gauss function, which remains
  the same (up to a scaling factor)  along the whole
  resonator. Hence the notions of transverse Hermite--Gauss modes and
  of Gaussian beams. (Gaussian beams also propagate outside the
  resonator.)

  \subsection{Stability}

So far, we assumed that orders of fractional Fourier transformations
associated with diffraction phenomena were real numbers.  By considering
Eq. (\ref{eq43}), we conclude that after a round trip, the field
amplitude is multiplied by a factor $\exp\,
(2\I\alpha_0)$ (the analysis of the phase factor will be done in Sect.
\ref{sect93}). The resonator is said to be stable \cite{Fog1}.

According to Eq. (\ref{eq39}), if
\begin{equation}
{(R'_1-L)(L+R_2)\over L(L-R'_1+R_2)}<0\,,\label{ineq54}\end{equation}
the order $\alpha_0$ is a complex number so that the factor $\exp
(2\I\alpha_0)$ represents an attenuation. There are losses by
diffraction and the resonator is said to be unstable (see part II of
the paper).

We conclude that a resonator is stable if, and only if, the order of
the fractional Fourier transformation associated with the field transfer
from a mirror to the other is a real number. 
If we change $R_2$ into $R'_2=-R_2$ in (\ref{ineq54}), then
a resonator of length $L$ is
stable if, and only if,
\begin{equation}
{(R'_1-L)(L-R'_2)\over L(L-R'_1-R'_2)}\ge
0\,,\label{ineq55}\end{equation}
where  $R'_1$ and $R'_2$ are the image radii of the mirrors. 
Inequality (\ref{ineq55}) is equivalent to
\begin{equation}
0\le \left(1- {L\over R'_1}\right)\left(1- {L\over R'_2}\right)\le
1\,,\end{equation}
which is a usual stability condition as given in most textbooks on
optical resonators \cite{Yar,Sie2}  

In the following, as previously stated, we consider only stable resonators.

\subsection{Consequence for the Wigner distribution of the field amplitude \\ inside a resonator}

First, we deduce  from Eq. (\ref{eq43}) that the
Wigner distribution associated with the scaled amplitude on ${\cal M}_1$
is invariant in a Wigner rotation of angle $-\alpha_0$. More
general, the analysis of Sect. \ref{sect73}
---Eq. (\ref{equ49})---shows that it is
invariant in every Wigner rotation. The result holds true for the scaled
field amplitude on every
appropriate wave surface inside or outside the resonator (by appropriate wave
surface, we mean a spherical cap whose radius is taken according
to Eq. (\ref{equ38})). We conclude that  the Wigner distribution associated with
  the scaled field amplitude inside a resonator is invariant in a Wigner
  rotation, whatever the  angle. The result also holds
  outside the resonator, that is, for Gaussian beams.

An equivalent statement is:

\smallskip
\noindent {\bf Theorem.} {\sl The Wigner distribution associated with
  the scaled field-amplitude of a transverse mode of  a resonator is invariant in
  a Wigner rotation, whatever the rotation angle.}

\smallskip
In the following, usual properties and usual relations that hold for stable
optical resonators and Gaussian beams are proved as a consequence of
the  previous theorem.

\subsection{Wigner distribution of a Gaussian field}

We consider a resonator as before. We assume the field amplitude on
  ${\cal M}_1$ to be of the form
\begin{equation}
U_1(\vec r)=U_0\,\exp\left(-{r^2\over {w_1}^{\! 2}}\right)\,,\label{eq53}\end{equation}
where $U_0$ is a dimensional constant, and $w_1$ ($w_1>0$) is called the
transverse radius of the field on ${\cal M}_1$, and is such that the emittance at
point $\vec r$ is higher than or equal to $|U_0|^2 \E^{-2}$ if, and only if, $r\le w_1$.

The Wigner distribution of the scaled field amplitude on ${\cal M}_1$ is
\begin{eqnarray}
W_1(\vec \rho,\vec \phi )\!\!\!\!&=&\!\!\!\! {\varepsilon_1R'_1\over \lambda}|U_0|^2
\int_{{\mathbb R}^2}\exp\left(-{\lambda\varepsilon_1R'_1\over
    {w_1}^{\! 2}}\left\Vert\vec\rho+{\vec\tau\over 2}\right\Vert^2\right)
\exp\left(-{\lambda\varepsilon_1R'_1\over
    {w_1}^{\! 2}}\left\Vert\vec\rho-{\vec\tau\over 2}\right\Vert^2\right)
\,\E^{2\I\pi\vec{\tau\cdot\phi}}\,\D\vec \tau \nonumber \\
&= &\!\!\!\!
{\varepsilon_1R'_1\over \lambda}|U_0|^2\,\exp\left(-{2\lambda
    \varepsilon_1R'_1\over {w_1}^{\! 2}}\rho^2\right)
\int_{{\mathbb R}^2}
\exp\left(-{\lambda
    \varepsilon_1R'_1\over 2{w_1}^{\! 2}}\tau^2\right)\,\E^{2\I\pi\vec{\tau\cdot\phi}}\,\D\vec \tau\,.
 \end{eqnarray}
We use (2--dimensional Fourier pair)
\begin{equation}
\exp\left(-{\lambda \varepsilon R'_1\over 2{w_1}^{\! 2}}\,{\tau^2}\right)\;\rightleftharpoons\,
{2\pi {w_1}^{\! 2}\over
    \lambda\varepsilon_1 R'_1}\,\exp\left(-{2\pi^2{w_1}^{\! 2}\over\lambda
    \varepsilon_1R'_1}\,\phi^2\right)\,,\end{equation}
and we obtain
\begin{eqnarray}
W_1(\vec \rho ,\vec \phi )&=&{2\pi {w_1}^{\! 2}\over \lambda^2}|U_0|^2
\exp \left(-{2\lambda \varepsilon_1 R'_1\over {w_1}^{\! 2}}\rho^2\right)
\exp\left(-{2\pi^2 {w_1}^{\! 2}\over \lambda \varepsilon_1 R'_1}\phi^2\right)\nonumber \\
&=&{2\pi {w_1}^{\! 2}\over \lambda^2}|U_0|^2
\exp \left(-{2\lambda \varepsilon_1 R'_1\over {w_1}^{\! 2}}\bigl({\rho_x}^2+{\rho_y}^2\bigr)\right)
\exp\left(-{2\pi^2 {w_1}^{\! 2}\over \lambda \varepsilon_1 R'_1}\bigl({\phi_x}^2+{\phi_y}^2\bigr)\right)
\,.\end{eqnarray}

\subsection{Transverse radius of the field  amplitude on a mirror}

We consider hyper-surfaces of equal amplitudes for the Wi\-g\-ner
distribution $W_1$ of the previous section. In the $(\vec \rho ,\vec
\phi )$--space, their equations are written
\begin{equation}
{\lambda \varepsilon_1R'_1\over {w_1}^{\! 2}}\rho^2+{\pi^2{w_1}^{\! 2}\over \lambda
  \varepsilon_1R'_1}\phi^2={\lambda \varepsilon_1R'_1\over {w_1}^{\! 2}}({\rho_x}^{\! 2}+{\rho_y}^{\! 2})+{\pi^2{w_1}^{\! 2}\over \lambda
\varepsilon_1R'_1}({\phi_x}^{\! 2}+{\phi_y}^{\! 2})=C\,,\label{eq57}\end{equation}
where $C$ is a constant.

The effect of diffraction on the Wigner distribution is a Wig\-ner rotation
that splits into two 2-rotations operating on planes $(\rho_x,\phi_x)$
and $(\rho_y,\phi_y)$ respectively. Then we consider equal amplitude sections of the
hyper-surface. 
These sections are curves
whose equations are written
\begin{equation}
{\lambda \varepsilon_1R'_1\over {w_1}^{\! 2}}{\rho_x}^{\! 2}+{\pi^2{w_1}^{\! 2}\over \lambda
  \varepsilon_1R'_1}{\phi_x}^{\! 2}=C_x\,,\label{eq57t}\end{equation}
in the $(\rho_x,\phi_x)$ plane, and
\begin{equation}
{\lambda \varepsilon_1R'_1\over {w_1}^{\! 2}}{\rho_y}^{\! 2}+{\pi^2{w_1}^{\! 2}\over \lambda
  \varepsilon_1R'_1}{\phi_y}^{\! 2}=C_y\,,\label{eq57q}\end{equation}
in the $(\rho_y,\phi_y)$ plane, where $C_x$ and $C_y$ are constants.

Generally, Eqs. (\ref{eq57t})  and (\ref{eq57q}) are  equations of
ellipses. But for eigenmodes of optical resonators, the curves of equal
amplitudes of the associated Wigner distribution are circles, because
they must be invariant by every Wigner rotation, according to the
previous theorem. 

Eq. (\ref{eq57t}) corresponds
to a circle if
\begin{equation}
{{w_1}^{\! 2}\over \lambda \varepsilon_1R'_1}= {\lambda \varepsilon_1R'_1\over
  \pi^2{w_1}^{\! 2}}\,.\label{eq58}\end{equation}
Then
\begin{equation}
{w_1}^{\!4}={1\over \pi^2}\lambda^2{\varepsilon_1}^{\! 2}{R'_1}^{\!2}\,.\label{eq66n}\end{equation}
From  Eqs. (\ref{eq3}) and (\ref{equ2}), and using $R_2=-R'_2$,  we obtain
\begin{equation}
{w_1}^{\!4}={\lambda^2 {R'_1}^{\!2} L (L-R'_2)\over \pi^2
  (R'_1-L)(L-R'_1-R'_2)}\,.\label{eq60}\end{equation}

Since propagation from ${\cal M}_1$ to ${\cal M}_2$ corresponds to a
rotation in the $(\rho_x ,\phi_x)$--plane, the above mentioned circle is
unchanged in the rotation, so that if $w_2$ is the field transverse radius on ${\cal M}_2$, we have
\begin{equation}
{{w_2}^{\! 2}\over \varepsilon_2R_2}={{w_1}^{\! 2}\over
  \varepsilon_1R'_1}\,,\label{eq61}\end{equation}
that is,
\begin{equation}
{w_2}^{\! 4}={\lambda^2 {R'_2}^{\! 2} L (L-R'_1)\over \pi^2
  (R'_2-L)(L-R'_1-R'_2)}\,.\label{eq62}\end{equation}
Eqs. (\ref{eq60}) and (\ref{eq62}) are classical relations for Gaussian
  beams in optical resonators \cite{PPF5}.

\subsection{Waist}\label{sect77}

We
consider the mirror ${\cal M}_1$ once more. 
We are looking for the field transverse radius $w$ on the spherical cap
${\cal S}$ (curvature radius $R$) at distance $D_1$ from ${\cal M}_1$. The field transfer from
${\cal M}_1$ to ${\cal S}$ is expressed by a fractional Fourier
transformation whose order is $\alpha$, and we denote $ \varepsilon$ the
parameter that corresponds to $\varepsilon_2$ in the general transfer
of Sect. \ref{sect2}. We know that equal amplitude curves of sections of the Wigner distribution are circles.
Then we adapt
Eqs. (\ref{eq58}) and (\ref{eq61}), and deduce
\begin{equation}
{w^2\over \lambda \varepsilon R}={{w_1}^{\! 2}\over \lambda
  \varepsilon_1R'_1}\,\cos^2\alpha +{\lambda \varepsilon_1R'_1\over
  \pi^2{w_1}^{\! 2}}\,\sin^2\alpha\,.\end{equation}
From Eqs. (\ref{equ2}) and (\ref{equ2b}) we obtain
\begin{equation}
w^2={w_1}^{\! 2} \,{R'_1-D_1\over R+D_1}\cdot{R\over
  R'_1}\cos^2\alpha
+{\lambda^2\over \pi^2{w_1}^{\! 2}}\cdot {{D_1}^{\!2} R'_1R\cot^2\alpha\over
  (R'_1-D_1)(R+D_1)}\,\sin^2\alpha\,.\label{eq65}\end{equation}
Finally, Eq. (\ref{eq3}) leads to
\begin{equation}
\cos^2\alpha ={(D_1+R)(R'_1-D_1)\over R'_1R}\,,\end{equation}
so that Eq. (\ref{eq65}) becomes
\begin{equation}
w^2={w_1}^{\!2}+{\lambda^2 {D_1}^{\!2}\over \pi^2{w_1}^2}+{w_1}^{\!2} {D_1\over R'_1}\left({D_1\over
  R'_1}-2\right)\,.\label{eq67}\end{equation}
Eq. (\ref{eq67}) provides the transverse radius of the field amplitude
  on an arbitrary wave surface ${\cal S}$ at a distance $D_1$
  from mirror ${\cal M}_1$. 

From Eq. (\ref{eq67}) we obtain $\D w^2/\D D_1=0$, if
\begin{equation}
D_1={R'_1\over 1+\displaystyle{\lambda^2{R'_1}^2\over
    \pi^2{w_1}^4}}\,.\label{eq77}\end{equation}
If we report  the value of $D_1$, as given by
Eq. (\ref{eq77}), in Eq. (\ref{eq67}) we obtain an extremum value of
$w^2$, which is ${w_0}^2$, given by
\begin{equation}
{w_0}^{\! 2}={{w_1}^{\! 2}\over 1+\displaystyle{\pi^2{w_1}^{\!4}\over \lambda^2{R'_1}^{\!2}}}\,.\label{eq78}\end{equation}
According to Eq. (\ref{eq67}), $w^2$ tends to $+\infty$ when
$D_1$ tends to $\pm \infty$. Then ${w_0}^{\! 2}$ is a minimum for $w^2$.

We now prove that the minimum $w_0$ ($w_0>0$) is obtained on a plane. First, we
use Eqs. (\ref{eq66n})  and (\ref{eq77}) and obtain
\begin{equation}
D_1={R'_1\over 1+\displaystyle{1\over {\varepsilon_1}^{\! 2}}}\,,\end{equation}
where $\varepsilon_1$ corresponds to the field transfer from ${\cal
  M}_1$ to ${\cal M}_2$. We use Eqs. (\ref{equ2}) and  (\ref{eq39}) so
that
\begin{equation}
D_1={L(L+R_2)\over 2L-R'_1+R_2}\,.\label{eq77t}
\end{equation}
We introduce then $D_2$, such that $L=D_1+D_2$, and obtain from Eq. (\ref{eq77t})
\begin{equation}
D_1(R'_1-L)+D_2(R_2+L)=0\,,\label{eq77n}\end{equation}
which means that the radius of ${\cal S}$ is infinite, according to
Eq. (\ref{equ38bis}).

We conclude that among the wave surfaces of the
$\{{\cal S}_\alpha\}$ family,
a surface exists, on which the  transverse radius  of the field is minimum. This
surface is a plane and its distance from ${\cal M}_1$ is given by
Eq. (\ref{eq77t}). The disk of points $\vec r$ such that $r\le w_0$  is the resonator waist. The
waist is on a plane and its transverse radius is $w_0$, as given by
Eq. (\ref{eq78}).

\subsection{The waist according to the resonator geometry}

We consider the previous resonator, with mirrors ${\cal M}_1$ and ${\cal M}_2$, and whose waist plane is denoted ${\cal W}_0$. We then consider an hypothetic  resonator whose mirrors are ${\cal M}_1$ and a plane
mirror located at ${\cal W}_0$, that is,
a resonator whose length is $D_1$, given by Eq. (\ref{eq77t}).
For infinite $R'_2$ (plane mirror), Eq. (\ref{eq62}) gives
\begin{equation}
{w_0}^{\!4}={\lambda^2\over \pi^2}D_1(R'_1-D_1)\,,\label{eq80}\end{equation}
and Eq. (\ref{eq77n}) leads to (we use $R'_2=-R_2$)
\begin{equation}
{w_0}^{\!4}={\lambda^2\over \pi^2} \, {L(R'_1-L)(L-R'_2)(L-R'_1-R'_2)
\over (2L-R'_1-R'_2)^2}\,,\label{eq76}\end{equation}
which gives the waist radius of the initial resonator (mirrors ${\cal
  M}_1$ and ${\cal M}_2$) as a function of its geometry. (Since
$\alpha_0$ is a real number, the numerator on the right side of
Eq. (\ref{eq76}) is positive according to Eq. (\ref{eq3}).)

\subsection{Rayleigh parameter}\label{sect81b}

The transverse radius of the field on a given  wave surface ${\cal S}$ of a Gaussian beam is denoted $w$, and we
  introduce the Rayleigh parameter on ${\cal S}$, that is, $\zeta$ such that
\begin{equation}
\zeta ={\pi\over \lambda}w^2\,.\end{equation}
The Rayleigh parameter is  $\zeta_0=\pi {w_0}^{\!2}/\lambda$ on the waist plane, and is known then
  as
  the Rayleigh distance.

We consider two spherical caps of the family $\{ {\cal S}_\alpha \}$, say
${\cal S}_{\alpha_1}$ (radius $R_1$) and ${\cal S}_{\alpha_2}$, with corresponding
Rayleigh parameters $\zeta_1$ and $\zeta_2$. From Eq. (\ref{eq67})
we obtain
\begin{equation}
\zeta_2=\zeta_1+{D^2\over \zeta_1}+\zeta_1{D\over R_1}\left({D\over
  R_1}-2\right)\,,\label{eq67t}\end{equation}
where $D$ is the distance (algebraic measure) from ${\cal S}_{\alpha_1}$ to ${\cal S}_{\alpha_2}$.

\subsection{Formulae related to the waist}\label{sect8a}

The waist is located on ${\cal S}_{\alpha_1}$, if this surface is a plane
  ($R_1$ is infinite).  Then $\zeta_1=\zeta_0$. According to
  Eq. (\ref{eq67t}), the Rayleigh
  parameter $\zeta$ on the surface ${\cal S}$ at distance $d$ from the
  waist is 
\begin{equation}
\zeta =\zeta_0+{d^2\over \zeta_0}\,.\label{eq68b}\end{equation}

If we set $D_1=-d$ and $R'_1=R$ in Eq. (\ref{eq80}), we obtain
the radius of curvature of the wave surface ${\cal S}$ at a distance $d$ from the
waist as
\begin{equation}R=-d-{\zeta_0^2\over d}\,.\label{eq82}\end{equation}

Equations (\ref{eq68b}) and (\ref{eq82}) are also classical in Gaussian
beam theory.

\section{The field amplitude on a wave surface. Longitudinal modes}\label{sect8}

\subsection{Fundamental mode. Gouy phase}
The results in the present section hold true for stable optical
resonators as well as for Gaussian beams.

The fundamental mode amplitude of an optical resonator is proportional
to a Gaussian function.
The amplitude on the waist
plane, chosen as reference for the phase,  is 
\begin{equation}
U(\vec r)=U_0\,\exp \left(-{r^2\over  {w_0}^{\!2}}\right)\,.\end{equation}

The field transfer from the waist to a wave surface ${\cal S}$ at a
distance $d$ is obtained by applying a fractional Fourier transform
whose order is $\alpha$, and according to Eq. (\ref{eq7}) diffraction
introduces a phase factor $\exp\, (\I\alpha
)$.

For infinite $R_1$, Eq. (\ref{eq3}) gives
\begin{equation}
\cot^2\alpha=-1-{R_2\over d}\,,\end{equation}
and then Eq. (\ref{eq82})
\begin{equation}
\tan^2\alpha = {d^2\over {\zeta_0}^{\!2}}\,.\end{equation}
Since $\alpha d >0$,  we have
\begin{equation}
\tan \alpha ={d\over \zeta_0}={\lambda d\over \pi {w_0}^{\!2}}\,.\label{eq86}\end{equation}

At last, we introduce the phase factor $\exp (-2\I\pi d/\lambda )$, which was not written in Eq. (\ref{eq1}).
If $w$ is the transverse radius on ${\cal S}$, the field amplitude on ${\cal S}$ is written
\begin{equation}
U_d(\vec r)= {w_0\over w} \,U_0\,\exp \left(-{r^2\over w^2}-{2\I \pi
    d\over\lambda}+\I\alpha\right)\,,\label{eq87}\end{equation}
where $\alpha$ is given by Eq. (\ref{eq86}). 

The factor $w_0/w$ in Eq. (\ref{eq87}) can be explained if we consider that the power
 that passes through  the waist (transverse radius $w_0$) is
also the power that passes through the circle of
radius $w$ on the wave surface ${\cal S}$. The power density is
  proportional to the square of
  the amplitude modulus and to the inverse of the area ($\pi {w_0}^{\!2}$ and $\pi
  w^2$).   Another explanation is based on the property of the
  Wigner distribution, that is, on Eq. (\ref{eq61}). If ${\cal S}_1$
    and ${\cal S}_2$ are two wave surfaces, then according to
      Eq. (\ref{eq61}), which is a consequence of the rotation invariance of
      the Wigner distribution of a Gaussian field, we obtain
\begin{equation}
{{w_1}^{\!2}\over
  {w_2}^{\!2}}={\varepsilon_1R_1\over\varepsilon_2R_2}\,.\end{equation}
From Eq. (\ref{eq28n}) and from the invariance of the Wigner distribution in
  a Wigner rotation,  we deduce
\begin{equation}
\int_{{\mathbb R}^2} \!\!|V_2|^2=\int_{{\mathbb R}^2} \!\!|V_1|^2\,,\end{equation}
and then, from  Eqs. (\ref{eq10}) and (\ref{eq11}),
\begin{equation}
{\displaystyle\int_{{\mathbb R}^2} \!\!\!|U_1|^2\over \displaystyle\int_{{\mathbb R}^2}\! \!\!|U_2|^2}= {\varepsilon_2R_2\displaystyle\int_{{\mathbb R}^2}\!\!\! |V_1|^2\over
  \varepsilon_1R_1\displaystyle\int_{{\mathbb R}^2}\!\!\! |V_2^2|}= {{w_2}^{\!2}\over {w_1}^{\!2}}\,.\label{eq93}\end{equation}
The factor $w_0/w$ in Eq. (\ref{eq87}) is obtained by applying Eq. (\ref{eq93}) to ${\cal
  S}$ and the waist plane. 

Finally, we remark that the order $\alpha$ in Eq. (\ref{eq87}) is usually
    called the Gouy phase \cite{Sie2} and is such that $\tan\alpha
    =d/\zeta_0$, according to Eq. (\ref{eq86}) \cite{Fog3}. We conclude that the effect of diffraction on
    Wigner distributions associated with eigenmodes of a resonator is a
    rotation whose angle is opposite to the Gouy phase. (The origin of
    the Gouy phase is taken on the waist plane.) 

\subsection{Higher modes}\label{sect92}
If $m$ and $n$ are two positive
 integers, the $(m,n)$--mode corresponds to the Hermite--Gauss function $\varphi_{m,n}$
defined by
 \begin{equation}
   \varphi_{m,n}(\xi,\eta)=H_m(\sqrt{2\pi}\,\xi)H_n(\sqrt{2\pi}\,\eta)\,\E^{-\pi
   (\xi^2+\eta^2)}\,,\end{equation}
where $H_m$ is the Hermite polynomial of order $m$, defined by
\begin{equation}
H_m(\xi)=(-1)^m\exp (\xi^2){\D^m\over \D  \xi^m}\exp (-\xi^2)\,.\end{equation}
Hermite--Gauss functions are eigenfunctions of
fractional order Fourier transforms, that is, for every $\alpha$ 
\begin{equation}
{\cal
  F}_\alpha[\varphi_{m,n}]=\exp [\I(m+n)\alpha]\,\varphi_{m,n}\,.\label{eq95}\end{equation} 
The field amplitude of the $(m,n)$--mode on the wave surface ${\cal
  S}$ at a distance $d$ from the waist is then
\begin{equation}
U_d(x,y)= {w_0\over w}\,U_0\, H_m\left(\sqrt{2}\,{x\over w}\right)\,
    H_n\left(\sqrt{2}\,{y\over w}\right) \, \exp \left[-{x^2+y^2\over w^2}-{2\I \pi
    d\over\lambda}+\I(1+m+n)\alpha\right]\,.\label{eq90}\end{equation}

\subsection{Resonator longitudinal modes}\label{sect93}

A factor $\exp \,(-2\I\pi D/\lambda )$ has been omitted in
Eq. (\ref{eq1}). We consider the resonator of Sect. \ref{sect72} once
more; its length is $L$, so that by taking into account the previous
factor, Eq. (\ref{eq43})  becomes
\begin{equation}
V_1=\exp \left(2\I\alpha_0-{4\I\pi L\over \lambda}\right){\cal F}_ {2\alpha_0}
[V_1]\,.\label{equ43n}\end{equation}

We assume that $V_1$ corresponds to the $(m,n)$--eigenmode, that is,
$V_1=\varphi_{m,n}$. We set $\alpha=\alpha_0$ in Eq. (\ref{eq95});  then
 Eq. (\ref{equ43n}) becomes 
\begin{equation}
V_{1}=\exp \left[2\I \alpha_0(1 +m+n)-{4\I\pi L\over \lambda}\right]
V_{1}\,.\label{equ43t}\end{equation}
We conclude that $L$, $\alpha_0$ and $\lambda$ are such that
\begin{equation}
{2\pi L\over \lambda}-\alpha_0(1+m+n)=q\pi\,,\label{eq101}\end{equation}
where $q$ is an integer.

Since $\alpha_0$ depends only on the resonator geometry (resonator
length and mirror radii), then only waves whose
wavelengths satisfy Eq. (\ref{eq101}) can propagate in the
resonator. Hence the notion of longitudinal modes.

\section{Concluding graphical analysis}\label{sect9}

We conclude with an elementary graphical analysis, which will be useful in discriminating stable and unstable
resonators (second part of the paper). We first change the order of variables: So far an arbitray
point in the phase space has been denoted $(\vec \rho, \vec
\phi)=(\rho_x,\rho_y,\phi_x,\phi_y)$; this point is now denoted $\vec
p=(\rho_x,\phi_x,\rho_y,\phi_y)=(P,Q)$ where $P$ and $Q$ are the
projections of $\vec p$ in the two-dimensional subspaces $(\rho_x,
\phi_x)$ and $(\rho_y,\phi_y)$. 
The Wigner distribution $W$ is changed into the function $W_s$ defined
by
\begin{equation}
W_s(\vec
p)=W_s(\rho_x,\phi_x,\rho_y,\phi_y)=W(\rho_x,\rho_y,\phi_x,\phi_y)\,.\end{equation}
The function $W_s$, as well as $W$, represents the field amplitude,
and its value $W_s(\vec p)$ represents the state of the field at the
scaled point
$(\rho_x,\rho_y)$ and 
scaled angular variables
$(\phi_x,\phi_y)$, that is,  after using  back scaled factors, the state
of the field at a point in the physical space and at a spatial
frequency. 

For the sake of simplicity we still call ``Wigner distribution'' the
function $W_s$, and from now on,  denote it by $W$.

\begin{figure}[h]
\begin{center}
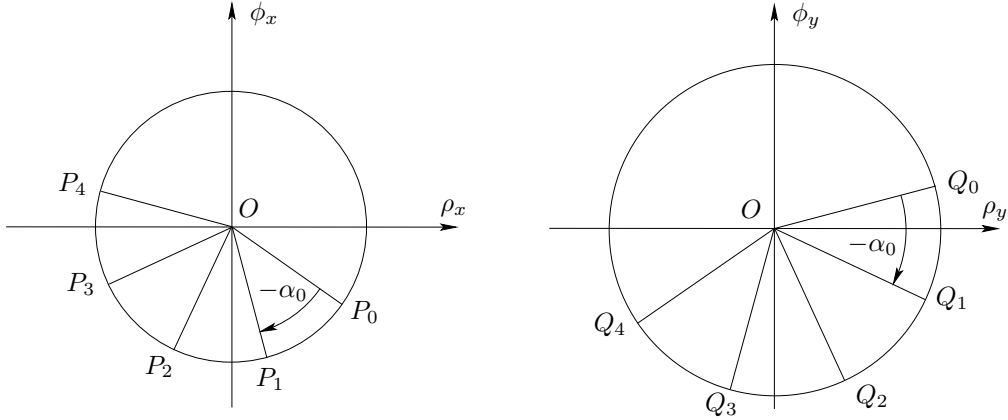
\caption{\small Graphical analysis of propagation between mirrors of a stable
  resonator. Planes $(\rho_x, \phi_x)$ and $(\rho_y,\phi_y)$ are
  section of the $4-$dimensional scaled phase-space. The value of the Wigner
  function is the same for every point $(P_j,Q_j)$ (and more generally
  for every point $(P,Q)$ where $P$ and $Q$ belong to the drawn circles). Point $P_{j+1}$
  (resp. $Q_{J+1}$) is deduced from $P_j$ (resp. $Q_j$) in a rotation of
  angle $-\alpha_0$. Distances $OP_j$
  and $OQ_j$ are constant, as a consequence of the resonator stability.\label{fig1}
}
\end{center}
\end{figure}

We consider a resonator made up of two mirrors ${\cal M}_1$ and ${\cal
  M}_2$ and denote $W_j$ the Wigner distribution associated with the
  field on ${\cal M}_j$. Let $\alpha_0$ be the order of the fractional
  Fourier transform associated with the field transfer from ${\cal
  M}_1$ to ${\cal
  M}_2$.
Let $\vec p_0$ be a point in the phase space. Then $W_1(\vec p_0)$ is also the value of the Wigner distribution
 $W_2$ associated with the optical field on mirror ${\cal M}_2$,  taken at
point $\vec p_1$ that is deduced from $\vec p_0$ in a 4--Wigner
rotation of angle $-\alpha_0$, that is  $W_2(\vec p_1)=W_1(\vec p_0)$. We have $\vec p_1=(P_1,Q_1)$, where $P_1$
(resp. $Q_1$)
is deduced from $P_0$ (resp. $Q_0$) in a 2--dimensional rotation of angle
$-\alpha_0$ (see Fig. \ref{fig1}).

We build a sequence of points $(\vec p_j)$ as follows: point
$P_{j+1}$ (resp. $Q_{j+1}$) is deduced from  $P_j$ (resp. $Q_j$) in the 2-rotation of angle
$-\alpha_0$ and  $\vec p_{j+1}=(P_{j+1},Q_{j+1})$. Let $d_j$ be the Euclidean distance
from $\vec p_j$ to the origin $O$. Then the sequence $(d_j)$ is a constant
sequence. This is also true for the distance from $P_j$ to $O$ and
from $Q_j$ to $O$ as shown in Fig. \ref{fig1}. This is equivalent to
saying that curves of constant amplitude of the Wigner distribution in
the 2--dimensional planes $(\rho_x, \phi_x)$ and $(\rho_y,\phi_y)$
are circles; or that the Wigner distribution
associated with a transverse mode of the field is invariant under Wigner
rotations, as seen before.

In the second part of the paper, we will consider diffraction
transfers that are represented by complex order fractional Fourier
transforms. We will prove that the effect on the Wigner distribution
reduces to two 2--dimensional hyperbolic rotations. We will also build
sequences of representative points, and will see that  sequences
$(d_j)$ are diverging and correspond to unstable resonators.


\section*{Appendix A. Proof of $\varepsilon_2R_2>0$}\label{appenA}
We assume $\alpha$ real and will  prove that $\varepsilon_1R_1$ and $\varepsilon_2R_2$ have the same
sign. We start
from the identity
$D(D-R_1+R_2)=(D-R_1)(D+R_2)+R_1R_2$, 
and deduce from Eq. (\ref{eq3}) and $\cot^2\alpha \ge 0$  ($\alpha$
real) that
\begin{equation}
{R_1R_2\over (R_1-D)(D+R_2)}\ge 1\,.\end{equation}
We conclude that
\begin{equation}
{R_1R_2D^2\over (R_1-D)(D+R_2)}\ge 0\,,\end{equation}
which means that $R_1D(R_1-D)$ and $R_2D(D+R_2)$ have the same
sign. Then from Eqs. (\ref{equ2}) and (\ref{equ2b}), we conclude that $\varepsilon_1R_1$ and $\varepsilon_2R_2$ have the
same sign.

\section*{Appendix B. Angle of rotation}\label{appenB}

Eq. (\ref{eq25}) corresponds to a rotation of the Wigner distribution, and
we  briefly explain  why  the angle of rotation is equal to $-\alpha$.


It will be enough to consider a function of two variables, say $f$.  Let $g$ be defined by
\begin{equation}
g(x,y)=f(x\cos\alpha -y\sin\alpha ,x\sin\alpha +y\cos\alpha)\,.\end{equation}
If $\alpha =\pi /2$, we have
$g(x,y)=f(-y,x)$, 
which means that the value of $g$ at point $P=(x,y)$ is the value of
$f$ at point $Q=(-y,x)$. We notice that $P$ is deduced from $Q$ in the
rotation of angle $-\pi /2$. We conclude that the graph of $g$ is
deduced from the graph of $f$ in a rotation of angle $-\pi
/2=-\alpha$. The result holds true for every $\alpha$.



\end{document}

%% file: fig0arxiv.pdf_t
\begin{picture}(0,0)%
\includegraphics{fig0arxiv.pdf}%
\end{picture}%
\setlength{\unitlength}{2279sp}%
\begingroup\makeatletter\ifx\SetFigFont\undefined%
\gdef\SetFigFont#1#2#3#4#5{%
  \reset@font\fontsize{#1}{#2pt}%
  \fontfamily{#3}\fontseries{#4}\fontshape{#5}%
  \selectfont}%
\fi\endgroup%
\begin{picture}(9564,4548)(154,-3682)
\put(2041,-2101){\makebox(0,0)[lb]{\smash{{\SetFigFont{8}{9.6}{\rmdefault}{\mddefault}{\updefault}{\color[rgb]{0,0,0}${\cal A}_1$}%
}}}}
\put(483,-631){\makebox(0,0)[b]{\smash{{\SetFigFont{8}{9.6}{\rmdefault}{\mddefault}{\updefault}{\color[rgb]{0,0,0}${C_1}$}%
}}}}
\put(2993,-830){\makebox(0,0)[rb]{\smash{{\SetFigFont{8}{9.6}{\rmdefault}{\mddefault}{\updefault}{\color[rgb]{0,0,0}$\vec r$}%
}}}}
\put(2944,-493){\makebox(0,0)[rb]{\smash{{\SetFigFont{8}{9.6}{\rmdefault}{\mddefault}{\updefault}{\color[rgb]{0,0,0}$M$}%
}}}}
\put(1171,-1816){\makebox(0,0)[rb]{\smash{{\SetFigFont{8}{9.6}{\rmdefault}{\mddefault}{\updefault}{\color[rgb]{0,0,0}$R_1$}%
}}}}
\put(9001,-2896){\makebox(0,0)[b]{\smash{{\SetFigFont{8}{9.6}{\rmdefault}{\mddefault}{\updefault}{\color[rgb]{0,0,0}${C_2}$}%
}}}}
\put(8101,-1771){\makebox(0,0)[lb]{\smash{{\SetFigFont{8}{9.6}{\rmdefault}{\mddefault}{\updefault}{\color[rgb]{0,0,0}$R_2$}%
}}}}
\put(5926,-2753){\makebox(0,0)[lb]{\smash{{\SetFigFont{8}{9.6}{\rmdefault}{\mddefault}{\updefault}{\color[rgb]{0,0,0}${\cal A}_2$}%
}}}}
\put(6286,-2281){\makebox(0,0)[b]{\smash{{\SetFigFont{8}{9.6}{\rmdefault}{\mddefault}{\updefault}{\color[rgb]{0,0,0}${\Omega_2}$}%
}}}}
\put(3316,-391){\makebox(0,0)[lb]{\smash{{\SetFigFont{8}{9.6}{\rmdefault}{\mddefault}{\updefault}{\color[rgb]{0,0,0}$m$}%
}}}}
\put(2776,112){\makebox(0,0)[lb]{\smash{{\SetFigFont{8}{9.6}{\rmdefault}{\mddefault}{\updefault}{\color[rgb]{0,0,0}$y$}%
}}}}
\put(3638,-722){\makebox(0,0)[lb]{\smash{{\SetFigFont{8}{9.6}{\rmdefault}{\mddefault}{\updefault}{\color[rgb]{0,0,0}$x$}%
}}}}
\put(6392,-751){\makebox(0,0)[lb]{\smash{{\SetFigFont{8}{9.6}{\rmdefault}{\mddefault}{\updefault}{\color[rgb]{0,0,0}$y'$}%
}}}}
\put(7068,-1455){\makebox(0,0)[lb]{\smash{{\SetFigFont{8}{9.6}{\rmdefault}{\mddefault}{\updefault}{\color[rgb]{0,0,0}$x'$}%
}}}}
\put(3488,-3106){\makebox(0,0)[b]{\smash{{\SetFigFont{8}{9.6}{\rmdefault}{\mddefault}{\updefault}{\color[rgb]{0,0,0}$D$}%
}}}}
\put(7095,-458){\makebox(0,0)[b]{\smash{{\SetFigFont{8}{9.6}{\rmdefault}{\mddefault}{\updefault}{\color[rgb]{0,0,0}${\cal P}_2$}%
}}}}
\put(3315,313){\makebox(0,0)[b]{\smash{{\SetFigFont{8}{9.6}{\rmdefault}{\mddefault}{\updefault}{\color[rgb]{0,0,0}${\cal P}_1$}%
}}}}
\put(2708,-1501){\makebox(0,0)[rb]{\smash{{\SetFigFont{8}{9.6}{\rmdefault}{\mddefault}{\updefault}{\color[rgb]{0,0,0}${\Omega_1}$}%
}}}}
\end{picture}%

%% file: fig1arxiv1bis.pdf_t
\begin{picture}(0,0)%
\includegraphics{fig1arxiv1bis.pdf}%
\end{picture}%
\setlength{\unitlength}{2486sp}%
\begingroup\makeatletter\ifx\SetFigFont\undefined%
\gdef\SetFigFont#1#2#3#4#5{%
  \reset@font\fontsize{#1}{#2pt}%
  \fontfamily{#3}\fontseries{#4}\fontshape{#5}%
  \selectfont}%
\fi\endgroup%
\begin{picture}(9931,4163)(877,-3301)
\put(5220,-1241){\makebox(0,0)[lb]{\smash{{\SetFigFont{10}{12.0}{\rmdefault}{\mddefault}{\updefault}{\color[rgb]{0,0,0}$\rho_x$}%
}}}}
\put(3320,669){\makebox(0,0)[lb]{\smash{{\SetFigFont{10}{12.0}{\rmdefault}{\mddefault}{\updefault}{\color[rgb]{0,0,0}$\phi_x$}%
}}}}
\put(3207,-1304){\makebox(0,0)[lb]{\smash{{\SetFigFont{10}{12.0}{\rmdefault}{\mddefault}{\updefault}{\color[rgb]{0,0,0}$O$}%
}}}}
\put(1779,-2042){\makebox(0,0)[rb]{\smash{{\SetFigFont{10}{12.0}{\rmdefault}{\mddefault}{\updefault}{\color[rgb]{0,0,0}$P_3$}%
}}}}
\put(1697,-1042){\makebox(0,0)[rb]{\smash{{\SetFigFont{10}{12.0}{\rmdefault}{\mddefault}{\updefault}{\color[rgb]{0,0,0}$P_4$}%
}}}}
\put(2561,-2861){\makebox(0,0)[rb]{\smash{{\SetFigFont{10}{12.0}{\rmdefault}{\mddefault}{\updefault}{\color[rgb]{0,0,0}$P_2$}%
}}}}
\put(3380,-2983){\makebox(0,0)[lb]{\smash{{\SetFigFont{10}{12.0}{\rmdefault}{\mddefault}{\updefault}{\color[rgb]{0,0,0}$P_1$}%
}}}}
\put(4320,-2318){\makebox(0,0)[lb]{\smash{{\SetFigFont{10}{12.0}{\rmdefault}{\mddefault}{\updefault}{\color[rgb]{0,0,0}$P_0$}%
}}}}
\put(3896,-2099){\makebox(0,0)[rb]{\smash{{\SetFigFont{10}{12.0}{\rmdefault}{\mddefault}{\updefault}{\color[rgb]{0,0,0}$-\alpha_0$}%
}}}}
\put(10622,-1256){\makebox(0,0)[lb]{\smash{{\SetFigFont{10}{12.0}{\rmdefault}{\mddefault}{\updefault}{\color[rgb]{0,0,0}$\rho_y$}%
}}}}
\put(8728,654){\makebox(0,0)[lb]{\smash{{\SetFigFont{10}{12.0}{\rmdefault}{\mddefault}{\updefault}{\color[rgb]{0,0,0}$\phi_y$}%
}}}}
\put(9775,-1638){\makebox(0,0)[rb]{\smash{{\SetFigFont{10}{12.0}{\rmdefault}{\mddefault}{\updefault}{\color[rgb]{0,0,0}$-\alpha_0$}%
}}}}
\put(10277,-1017){\makebox(0,0)[lb]{\smash{{\SetFigFont{10}{12.0}{\rmdefault}{\mddefault}{\updefault}{\color[rgb]{0,0,0}$Q_0$}%
}}}}
\put(10160,-2176){\makebox(0,0)[lb]{\smash{{\SetFigFont{10}{12.0}{\rmdefault}{\mddefault}{\updefault}{\color[rgb]{0,0,0}$Q_1$}%
}}}}
\put(8425,-1308){\makebox(0,0)[rb]{\smash{{\SetFigFont{10}{12.0}{\rmdefault}{\mddefault}{\updefault}{\color[rgb]{0,0,0}$O$}%
}}}}
\put(9689,-3159){\makebox(0,0)[rb]{\smash{{\SetFigFont{10}{12.0}{\rmdefault}{\mddefault}{\updefault}{\color[rgb]{0,0,0}$Q_2$}%
}}}}
\put(8186,-3237){\makebox(0,0)[rb]{\smash{{\SetFigFont{10}{12.0}{\rmdefault}{\mddefault}{\updefault}{\color[rgb]{0,0,0}$Q_3$}%
}}}}
\put(7075,-2437){\makebox(0,0)[rb]{\smash{{\SetFigFont{10}{12.0}{\rmdefault}{\mddefault}{\updefault}{\color[rgb]{0,0,0}$Q_4$}%
}}}}
\end{picture}%

%% file: arxiv1V2.bbl
\begin{thebibliography}{references}\label{refe}


\leftskip = -1cm



 \bibitem{Alo}  M. A. Alonso, ``Wigner functions in optics: describing
beams as ray bundles and pulses as
particle ensembles'',
Advances in Optics and Photonics, {\bf 3}
  (2011) 272--365.


\bibitem{Wal} A. Walther,  ``Radiometry and Coherence'', J. Opt. Soc. Am. {\bf 58} (1968) 1256--1259.

\bibitem{Bas} M. J. Bastiaans, ``Wigner distribution function and its application to first-order optics'', J. Opt. Soc. Am. {\bf 69} (1979)
  1710--1716.

\bibitem{Cuy} T. Cuypers, T. Haber, P. Bekaert, S. B. Oh, R. Raskar, ``Reflectance Model for Diffraction'',
  ACM Transactions on Graphics, {\bf 28} (2009) Art. 106.

 \bibitem{Mou} B. M. Mout, M. Wick, F. Bociort, H. P. Urbach, ``A
  Wigner-based-ray-tracing method for imaging simulations'', in {\em
  Optical Systems Design: Computational Optics}, Daniel G. Smith,
  Frank Wyrowski, Andreas Erdmann Eds., {\em Proc. SPIE} {\bf 9630} (2015)
  96300Z-1--96300Z-11.
 
\bibitem{Men} D. Mendlovic, H. M. Ozaktas, A. W. Lohmann, ``Graded-index fibers, Wigner-distribution
functions, and the fractional Fourier transform'',
  Appl. Opt. {\bf 33} (1994) 6188--6193.

\bibitem{Oza3} H. M. Ozaktas, Z. Zalevsky, M. A. Kutay,  {\it The fractional
    Fourier transform with applications in optics and signal processing}, John
    Wiley \& Sons, Chichester, 2001.

\bibitem{Loh} A. W. Lohmann,  ``Image rotation, Wigner rotation and the
  fractional Fourier transform'', J.~Opt. Soc. Am. A {\bf 10} (1993)
  2181--2186.

\bibitem{Coe1} S. Co\"etmellec, D. Lebrun, C. \"Ozkul,
  ``Characterization of diffraction patterns directly from in-line
  holograms with the fractional Fourier transform'', Appl. Opt. {\bf 41} (2002) 312--319.

\bibitem{Tes} M. Testorf, ``The phase--space approach to optical system theory'',
 Optics and Photonics Letters {\bf 5} (2013)
  1330001.

\bibitem{PPF1}
 {P. Pellat-Finet},  ``Fresnel diffraction and the fractional order
 Fou\-rier transform'', Opt.  Lett. {\bf 19} (1994) 1388--1390.

\bibitem{PPF3}
P. Pellat-Finet, G. Bonnet, ``Fractional order Fourier transform and
Fourier optics'',  Opt. Comm. {\bf 111 }
(1994) 141--154.

\bibitem{PPF5} P. Pellat-Finet, {\sl Optique de Fourier. Th\'eorie m\'etaxiale
  et fractionnaire}, Springer, Paris, 2009.

\bibitem{Alm} L. B. Almeida, ``The fractional Fourier transform and
  time-frequency representations'', IEEE Trans. Sign. Process., {\bf 42} (1994) 3084--3091.

\bibitem{Rom} C. J. Rom\'an-Moreno, R. Ortega-Mart\'inez,
  C. Florez-Arviso, ``The Wigner function in paraxial optics II. Optical diffraction pattern
                         representation'', Revista Mexicana de F\'isica {\bf 49} (2003) 290--295.

\bibitem{Nam} {V. Namias}, ``The fractional order Fourier transform and
  its applications to quantum mechanics'', J. Inst. Maths Applics {\bf 25}
(1980) 241--265.

\bibitem{PPF4}  P. Pellat-Finet, P.-E. Durand, ``La notion de spectre
    angulaire sph\'erique'', C. R. Physique  {\bf
    7} (2006) 457--463.

\bibitem{PPF4b}  P. Pellat-Finet, P.-E. Durand, \'E. Fogret, ``Spherical
    angular spectrum and the fractional order Fourier transform'', Opt. Lett.  {\bf
    31} (2006) 3429--3431.

\bibitem{Goo} J. W. Goodman, {\sl Introduction to Fourier optics},
  3$^{\rm d}$ Ed., Robert \& Company, Englewood,
  2005.

\bibitem{Fog1}  P. Pellat-Finet, \'E. Fogret,  ``Complex order fractional
  Fourier transforms and their use in diffraction theory'', Opt. Comm. {\bf 258}
(2006) 103--113.

\bibitem{Fog2}  \'E. Fogret, P. Pellat-Finet, ``Agreement of fractional
  Fourier optics with the Huygens--Fresnel principle'', Opt. Comm. {\bf 272}
  (2007) 281--288.

\bibitem{Fog3} {P. Pellat-Finet, \'E. Fogret}, ``Fractional Fourier optics theory of
  optical resonators'', in: P. S. Emersone (Ed.), {\sl Progress in optical fibers}, Nova
  Science Publishers, New York (2011) 299--351.

  \bibitem{Yar} {A. Yariv, P. Yeh}, {\sl Photonics. Optical electronics in modern communications}, 6$^{\rm th}$ Ed., Oxford University Press, New York, 2007.


\bibitem{Sie2} {A. E. Siegman.}, {\sl Lasers}, University Science Books, Mill Valley, 1986.



\end{thebibliography}
